\documentstyle[12pt]{article}

\begin{document}
\title{Quantum field theory: Finiteness and Effectiveness\footnote{
This is an extended version of the invited talk presented at the 11th
International Conference (PQFT'98) held in JINR, Dubna, Russia, 
July 13-17, 1998.}}
\author{Jifeng Yang\\
 School of Management \\
Fudan University, Shanghai, 200433, P. R. China}
\date{\today}
\maketitle
\begin{abstract}
A new attempt is demonstrated that QFTs can be UV finite if 
they are viewed as the low energy effective theories of a 
fundamental underlying theory (that is complete and well-defined 
in all respects) according to the modern standard point of 
view. This approach works for any interaction model and space-time 
dimension. It is much simpler in principle and in technology 
comparing to any known renormalization program.Unlike the 
known renormalization methods, 
the importance of the procedure for defining the ambiguities 
(corresponding to the choice of the renormalization conditions 
in the conventional program) is fully appreciated in the new 
approach. It is shown that the high energy theory(s) or the 
underlying theory(s) in fact 'stipulates (stipulate)' the low 
energy and effective ones through these definitions within our 
approach while all the conventional methods miss this important 
point. Some simple but important nonperturbative examples are 
discussed to show the power and plausibility of the new 
approach.Other related issues ( especially the IR problem 
and the implication of our new approach for the canonical 
quantization procedure) are briefly touched.
\end{abstract}

PACS number(s): 11.10.-z; 11.10.Gh; 11.15.-q; 11.15. Bt
\newpage
\section{Introduction}

It is known to all that the old frameworks of renormalization 
(Ren) first invoke UV infinities and then try to find some doubtful 
'operation' to remove them in order to predict the obviously finite
world \cite {Jac}. The worse is, one has to find a regularization 
(Reg) first in the intermediate stage of the framework without 
appreciating the physical implication of this technical necessity.
In short, the difficulty is inevitable if one holds the present 
formulation of QFTs to be complete and elementary. The necessity 
of introducing a regularization (in whatever way \cite {DR}) 
itself means already that the present formulation of QFTs is not 
a complete or fundamental one. This is also reflected in that
most field theorists take the QFTs as defined below a UV cutoff 
scale.
But such a cutoff scenario often pushes us to the difficulties
like removing infinities in ways as consistent as possible.

Now, it has become a standard point of view that a 
fundamental theory (well defined for the extremely high energy 
end) underlies the present QFTs that are in fact low energy 
(LE) effective theories
for the phenomena in LE ranges \cite {wein}. But as far as 
the author knows, we are still lacking a formulation that can 
yield finite results in a natural way (without invoking {\it ad 
hoc } Regs and divergences) that fully makes use of the
standard point of view. A new approach is proposed in 
Ref. \cite {YYY} that fully exhibits the power of the standard 
point of view if one uses it appropriately. (The Wilsonian 
approach \cite {wil} which works perfectly in the context of 
critical phenomena, is questionable if one applies it to all 
ordinary QFTs in the original sense as then it can only deal 
with the renormalizable ones in an {\it ad hoc } way (see, 
Eq.(18) in Ref. \cite {pol}, which needs verification instead 
of being imposed for the theories).) As will be seen in the 
following, our approach is rather simple and does not 
depend on model specifics and space-time dimension. 
(One may take it as a strategy more than as a technology 
due to its wide applicability.)

Let us elaborate on the standard point of view as a natural 
postulate or argument: suppose, the true complete theory 
underlying the present QFTs is found, {\it it must be well 
defined in every aspect and always yields physically 
sound (finite, of course) predictions in any energy range, at 
least for those ranges supposed to be well described by 
present QFT models. It must have been characterized by 
certain new parameters dominant in the extremely high 
energy end to make the theory UV finite. } All the 
objects (like the FAs) given by the present formulation 
of QFTs should be well-defined if they are derived or
calculated from the underlying theory with certain limit 
operation about its fundamental parameters performed 
afterwards as we are presently in a "low energy" phase. 
(we do not view FAs as primary starting points but as 
something derived from the underlying theory.) Of course 
these Green functions (calculated from the underlying theory) 
for the LE phenomena 
compromise only various subsets of Green functions (in the 
underlying theory). Our present QFTs are just LE
{\it reformulations or reorganizations } via the present 
quantization procedures without the information about the UV 
underlying world (hence being possibly UV ill-defined ones).

In some treatments, one simply 'integrates out' the 'HE' modes 
of a presently known model. This is not correct as we will see 
below. In fact, the mechanism for the appearance or evolution 
of LE fields and phenomena from the underlying theory 
must be a sophisticated 'integrating-out'  of the modes given 
by the underlying theory rather than those described by the 
present (LE effective) models. When we are talking about the 
'integrating-out' of HE modes or the underlying 
modes, we refer to the mechanisms--unknown to us yet--of 
the emergence or 'evolution' of the LE phenomenological 
modes or fields out of the underlying theory.

To focus on the UV problem, we will assume from now on 
that there were no unphysical IR singularity in the LE models 
in our discussions or we have already had an IR regular 
formulation for the LE QFTs. (We will discuss later about the 
IR structure's contribution to the whole formulation--it should 
be reconsidered to arrive at a totally satisfying formulation, 
esp. for QCD-like theories where the IR singularity is rather 
serious and affects the theories' predictions \cite {Bigi}).
 
It is desirable to employ a generating functional formalism 
\cite {Shirkov} or a path integral formalism to assemble 
these Green functions for each of the subsets. It is natural 
to expect that generally the well-defined path integral should 
necessarily carry some indispensable information about the
underlying theory, i.e., the fundamental parameters
characterizing the underlying theory, in the following way
\begin{equation}
Z_{\{\sigma\}}(\{J^{i}\})=
\int D{\mu}(\phi^{i}_{\{\sigma\}}) 
\exp \{iS(\phi^{i}_{\{\sigma\}};\{J_{i}\}; \{\sigma\})\}
\end{equation}
where $\{\sigma\}$ are the underlying fundamental parameters 
(some fundamental constants which should include the 
Newtonian gravitation constant) from the underlying theory and 
$\{J^{i}\}$ are the external sources specifying the LE 
phenomenon. The 'elementary fields' for the QFTs 
$(\phi^{i}_{\{\sigma\}})$ 
are here appended by the underlying  parameters to indicate 
that they are in fact effective. It is easy to see that for 
different LE phenomenological physics, the LE limit operation 
may act upon sets of underlying parameters that differ in part. 

The ease of using the path integral formalism lies in that one can
again think in terms of Hamiltonian or Lagrangian which is more
familiar to physicists. Thus it is immediate to see that the spectra
given by the conventional field theories' Hamiltonians differ from 
those derived from the underlying theory, especially in the UV 
regions. And the deviation is in a sense measured by the 
ill-definedness or UV divergence in the conventional QFTs. More 
severe UV divergence implies more degrees of deviation, as is 
expressed through the fact that unrenormalizable models which 
exhibit more serious divergence are in fact models suitable for 
much lower energy ranges, i.e., these models are worse
descriptions for higher energy physics. The underlying theory 
picture means that these effective fields (or modes) will break 
up when energy goes up, in the meantime new and more 
elementary field or modes become active. Of course some of
the fields or modes might persist over all high energy ranges, 
which means that they were elementary modes in the 
underlying theory (cosmic 'fossils' of the big Bang?). Anyway, 
the true spectral manifolds of fields (which must be more 
complicated ones in terms of the underlying parameters)
in the LE phenomena are not the simple ones (often Euclidean 
spaces in the UV ends) given by the present Hamiltonians. 
Hence, it is not justified to extrapolate linearly the spectra 
probed in the LE ranges ( and so the simple 'integrating-out' 
of the 'HE' modes given by any LE model is not justified).

The true spectral manifolds are presently beyond our reach. 
Both the underlying parameters $\{\sigma\}$ and the way they 
enter into the 'story' are unknown yet. Without them, one can 
not calculate anything in principle from our present QFTs due 
to the ill-definedness. The usual way is to cut off the UV parts 
of the spectra by hand in order to be able to compute. That is
the historical origin for the regularization (Reg)procedures. But 
this artificial operation often 'deforms' the effective theories
(without physical justification) in such a way that when these 
deformations are removed, infinite results may appear. 
This in turn calls for the procedures of subtraction or Rens. 
All these troubles are due to not knowing how the underlying 
parameters work. The Regs are just artificial substitutes for the 
$\{\sigma\}$.

What we are trying to present in the following is that if one 
starts merely with the existence of the underlying theory and 
$\{\sigma\}$ (without knowing the details), then there is a 
simple way one can calculate the amplitudes wanted 
without introducing any {\it ad hoc } Reg or cutoff that leads 
to UV divergence.
 
The principle and core technical part of our approach are given 
in section II. We first demonstrate it for the one-loop case in the
Feynman graph language. The treatment of the multi-loop cases 
is given in section III where many conventional subtleties like
overlapping divergence and shifting of integral momenta are 
shown to be easily resolved. Some general issues associated 
with the whole structures of QFTs are given there. Then we 
discuss some nonperturbative examples in our approach in 
section IV where some quantum mechanical cases are shown 
to be in fact supporting our point of view concerned with some 
weak points of the old Ren framework. Section V is devoted to 
the discussion of the IR problem, which points out some 
directions for further investigation and its nontrivial relation to 
the whole theory structures. The last 
section contains some discussions and the summary.

\section{How Can UV Finite Results be Derived}

From our discussion in section I, we see that the effective
Hamiltonians (and hence the propagators and vertices in their 
present forms) are the LE limits of the ones characterized by 
$\{\sigma\}$. In an ill-defined Feynman amplitude (or other 
quantities in different formalisms) constructed from the present 
forms of propagators and vertices, we had in fact first taken the 
LE limit operation before any internal integration is done. Thus, 
it is immediate to observe that: {\it ill-defined (or divergent) 
Feynman amplitudes (FAs) (or quantities in other formulation) 
given by the effective theories (QFTs) are consequences 
of illegitimate operations on the corresponding "amplitudes" 
from the underlying theory. } In formula, if  the integrand  
$ f(\{Q_{i}\}, \{p_{j}\},\{m_{k}\})$ of an ill-defined FA corresponds 
to the integrand $\bar{f} (\{Q_{i}\},\{p_{j}\},\{m_{k}\}; \{{\sigma}_{l}\})$ 
from the underlying theory with 
$\{Q_{i}\}, \{p_{j}\},\{m_{k}\},\{{\sigma}_{l}\}$ being 
respectively loop momenta, external momenta, masses and the 
fundamental parameters in the underlying theory, then
\begin{eqnarray}
&\Gamma^{0}&(\{p_{j}\},\{m_{k}\}) = {\bf L}_{\{\sigma\}}
\overline {\Gamma} (\{p_{j}\},\{m_{k}\};\{\sigma_{l}\})\nonumber \\
&= &{\bf L}_{\{\sigma\}} \int \prod_{i} d^{n}Q_{i} \bar{f}
(\{Q_{i}\},\{p_{j}\},\{m_{k}\};\{\sigma_{l}\})\nonumber \\
&\neq& \int \prod_{i}d^{n} Q_{i} {\bf L}_{\{\sigma\}} \bar{f}
 (\{Q_{i}\},\{p_{j}\},\{m_{k}\};\{\sigma_{l}\})\nonumber \\
&=& \int \prod_{i} d^{n}Q_{i} f(\{Q_{i}\},\{p_{j}\},\{m_{k}\}),
\end{eqnarray}
where $\Gamma^{0}$ and $\overline {\Gamma}$ are well-defined 
(finite), the symbol ${\bf  L}_{\{\sigma\}}$ denotes the LE limit 
operations and $n$ refers to space-time dimension. That means, 
${\bf L}_{\{\sigma\}}$ and $\int \prod_{i} d^{n}Q_{i}$ do not
commute on all the integrands $\bar{f}(...)$, i.e., the commutator 
\begin{equation}
\delta_{\{\sigma\}}= \left [ {\bf L}_{\{\sigma\}},
\int \prod_{i} d^{n}Q_{i} \right ]
\end{equation}
only vanishes identically for convergent (i.e., well-defined) FAs, 
otherwise we meet troubles: divergence or ill-definedness in FAs. 
That is to say, the deviation of the effective formalism is not 
detected by the convergent FAs, or these amplitudes can be 
well described by the LE limit forms of the effective theories. This 
is an extremely important fact for our purpose in the following.

As the underlying theory or the amplitudes 
$\bar{f}(...;\{\sigma_{l}\})$ are unavailable by now, we have to 
find a way to approach the truth, 
$\Gamma^{0}(\{p_{j}\},\{m_{k}\})$'s. In the following, 
we will demonstrate a new and tractable way to achieve this 
goal which is different from any existent methods (see, e.g. 
Ref \cite {pol} that are based on the Wilson's picture \cite {wil}. 
We will discuss later why the old ones \cite {pol} are in fact 
rather limited and {\it ad hoc } and simply incapable of 
dealing with the 'unrenormalizable' interactions that are 
physically interested).

First we show that the following important relation holds for 
1-loop case ill-defined FAs (c.f. Eq.(2) for 1-loop case)
\begin{equation}
\int d^{n}Q \left ({\partial}_{p_{j}} \right )^{\omega} 
f(Q,\{p_{j}\},\{m_{k}\})= 
\left ( {\partial}_{p_{j}} \right )^{\omega} \Gamma^{0} 
(\{p_{j}\},\{m_{k}\}),
\end{equation}
with $\omega-1$ being the usual superficial divergence degree 
of $\int d^{n}Q f (Q,\{p_{j}\},\{m_{k}\})$ so that the lhs of Eq.(4) 
exists (finite), $\left ({\partial}_{p_{j}} \right )^{\omega} $ 
denoting differentiation's wrt the external parameters 
$\{p_{j}\}$'s of the amplitude and $\Gamma^{0}(...)$ is the LE
limit of the amplitude calculated in the underlying theory (i.e.,
the internal momentum integration is performed first). It is easy to
see that the operation $\left ({\partial}_{p_{j}} \right )^{\omega} $
leads to convergent graphs with certain external momenta
taking zero values at the new vertices hence "created"  and the
convergent graphs can be calculated with the present forms of 
propagators and vertices or with performing the LE limit operation
first as it now commutes with the internal momentum integration.

The proof is very simple, since
\begin{eqnarray}
&\int& d^{n}Q  \left ({\partial}_{p_{j}} \right )^{\omega} 
f (Q,\{p_{j}\},\{m_{k}\})= \int d^{n}Q  
\left ({\partial}_{p_{j}} \right )^{\omega} {\bf  L}_{\{\sigma\}} 
\bar{f} (Q,\{p_{j}\},\{m_{k}\};\{\sigma_{l}\})\nonumber \\
&=& \int d^{n}Q {\bf  L}_{\{\sigma\}} \left ({\partial}_{p_{j}} 
\right )^{\omega}\bar{f} 
(Q,\{p_{j}\},\{m_{k}\};\{\sigma_{l}\})\nonumber \\
&=& {\bf  L}_{\{\sigma\}}  \int d^{n}Q 
\left ({\partial}_{p_{j}} \right )^{\omega}\bar{f}
(Q,\{p_{j}\},\{m_{k}\};\{\sigma_{l}\})\nonumber \\
&=&{\bf L}_{\{\sigma\}} \left ({\partial}_{p_{j}} \right )^{\omega}
\overline{\Gamma} (\{p_{j}\},\{m_{k}\};\{\sigma_{l}\}) = 
\left ({\partial}_{p_{j}} \right )^{\omega} \Gamma^{0}
 (\{p_{j}\},\{m_{k}\}).
\end{eqnarray}
The second and the fifth steps follow from the commutativity of  
the two operations $\left ({\partial}_{p_{j}} \right )^{\omega}$ and 
${\bf  L}_{\{\sigma\}}$ as they act on different arguments, the third 
step is due to the existence of $\int d^{n}Q \left ({\partial}_{p_{j}}
\right )^{\omega} f (Q,...)$ and the fourth is justified from the 
existence of $\int d^{n}Q \bar{f} (Q,...;\{\sigma_{l}\})
( = \overline {\Gamma} (...;\{\sigma_{l}\}))$.

The right end of Eq.(4) can be found now as the left end exists 
as a nonpolynomial (nonlocal) function of external momenta 
and masses, i.e., denoting it as $\Gamma^{0}_{(\omega)}$, 
\begin{equation}
\left ({\partial}_{p_{j}}\right )^{\omega} \Gamma^{0} 
(\{p_{j}\},\{m_{k}\}) = \Gamma^{0}_{(\omega)} 
(\{p_{j}\},\{m_{k}\}).
\end{equation}
To find $\Gamma^{0} (\{p_{j}\},\{m_{k}\})$, we integrate both 
sides of Eq.(6) wrt the external momenta "$\omega$" times 
indefinitely to arrive at the following expressions
\begin{eqnarray}
& &\left (\int_{{p}}\right )^{\omega}
 \left [ ({\partial}_{{p}})^{\omega} \Gamma^{0} 
(\{p_{j}\},\{m_{k}\}) \right ] = \Gamma^{0} (\{p_{j}\},\{m_{k}\}) 
 +  N^{\omega} (\{p_{j}\},\{c_{\omega}\}) \nonumber \\
&=& \Gamma_{npl} (\{p_{j}\},\{m_{k}\}) + N^{\omega} 
(\{p_{j}\}, \{C_{\omega}\}) 
\end{eqnarray}
with $\{c_{\omega}\}$ and $\{C_{\omega}\}$ being arbitrary 
constant coefficients of an $\omega-1$ order polynomial in 
external momenta $N^{\omega}$ and 
$\Gamma_{npl} (\{p_{j}\},\{m_{k}\})$ being a definite 
nonpolynomial function of momenta and masses \cite {JF}. 
Evidently $\Gamma^{0} (\{p_{j}\},\{m_{k}\}) $ is not uniquely 
determined within conventional QFTs at this stage. That the 
true expression
\begin{equation}
 \Gamma^{0} (\{p_{j}\},\{m_{k}\}) = \Gamma_{npl} 
(\{p_{j}\},\{m_{k}\}) + N^{\omega} 
(\{p_{j}\},\{\bar{c}_{\omega}\}) , \ \ \ \bar{c}_{\omega}= 
C_{\omega}-c_{\omega}
\end{equation}
contains a definite polynomial part (unknown yet) implies that it 
should come from the LE limit operation on 
$\overline {\Gamma} (\{p_{j}\},\{m_{k}\};\{\sigma_{l}\})$ (see Eq.(2)) 
as the usual convolution integration can not yield a polynomial 
part--also an indication of the incompleteness of the formalism 
of the QFTs.

We can take the above procedures as efforts for rectifying the 
ill-defined FAs and "represent" the FAs with the expressions 
like the rhs of Eq.(7), i.e., 
\begin{equation}
\int d^{n}Q f (Q,\{p_{j}\},\{m_{k}\}) >=< \Gamma_{npl} 
(\{p_{j}\},\{m_{k}\}) + N^{\omega} (\{p_{j}\}, \{C_{\omega}\})
\end{equation}
with "$>=<$" indicating that lhs is rectified as rhs \cite {JF}. That 
the ambiguities reside only in the local part means that the 
QFTs are also quite effective over a nonzero space-time distance.

To find the $ \{\bar{c}_{\omega}\}$'s in Eq.(8) we need inputs from 
the physical properties of the system ( such as symmetries, 
invariances, unitarity of scattering matrix and reasonable behavior 
of differential cross-sections) and a complete set of  data from 
experiments \cite {CK,LL} (if we can derive them from the 
underlying theory all these requirements would be automatically 
fulfilled) as {\sl physics determines everything after all. }  In other 
words, all the ambiguities should be fixed in this way. Note that 
this is a principle independent of interaction models and space-time 
dimensions, i.e., we can calculate the quantum corrections in any 
model provided the definitions can be consistently and effectively 
done. Similar approach had been adopted by Llewellyn Smith to 
fix ambiguities on Lagrangian level by imposing high energy 
symmetry, etc. on relevant quantities \cite {LL}. 

For later use, I would like to elaborate on the implications 
of the constants. As we have seen, the $\bar{c}_{\omega} $'s 
arise in fact from the low energy limit operation on the objects 
already calculated in the underlying theory, they are uniquely 
defined for any specific low energy phenomenology up to possible 
reparametrization invariance. Different choices of these 
constants either are incorrect or simply correspond to different
LE theories (amount to being defined by different underlying
theories). Since different Regs and/or Ren conditions 
correspond to different choices of the constants,
we may find, especially in nonperturbative cases (we will see 
such important examples in section IV that support our 
arguments here \cite {QMDR}), that different Regs and Ren 
conditions lead to rather different 'renormalized' LE theories, 
or even could not describe relevant low energy physics. Thus 
it is clear that the low energy effective theories can not be 
totally independent of the underlying theory(s), i.e., 
{\it the underlying theory stipulates or influences the effective 
ones through these constants though the fundamental 
parameters characterizing the underlying theory do not 
appear in the LE formulations. } All the known approaches 
seemed to have failed to fully appreciate this important part. 
As we will see in section IV, the former studies on the 
self-adjoint extension \cite {QM} of some quantum 
mechanical Hamiltonians just confirm our conclusions here.

\section{Multi-loop Case}

The treatment for the multiloop case is very simple and 
straightforward, at least in principle.

Since the UV divergence will appear if one first take the limit
before doing loop momenta integrations, our strategy is just 
to move the limit operator ${\bf L}_{\{\sigma\}}$ across the 
integration operations in such a way that no potential 
divergence is left over just like in the single loop case.

For any multi-loop graph $\Gamma$ (we will use the same 
symbol to denote the graph and the FA associated with if it is 
not confusing), we should start with the corresponding 
amplitude derived from the 
underlying theory, i.e.,$\overline {\Gamma}(\ldots;\{\sigma\})$ 
with the same graph structure. The only difference, as with the 
1-loop case, lies in that all the internal lines and vertices are 
understood to be given by the underlying theory, which are 
necessarily characterized by the presence of the parameters 
$\{\sigma\}$ and are expressed in forms unknown to us yet.
For our purpose, it is enough to know they exist. Then the LE 
limit of $\overline{\Gamma}(\ldots;\{\sigma\})$ is just (denoted 
as $\Gamma^0(\ldots;\{c^0\})$ with $\{c^0\}$ indicating the 
definite constants unknown to us that are left over by the LE 
limit operation)
\begin{eqnarray}
\Gamma^0(\ldots;\{c^0\})
&=& {\bf L}_{\{\sigma\}} \overline{\Gamma}(\ldots;\{\sigma\}) 
       \nonumber \\
&=& {\bf L}_{\{\sigma\}} \int \prod_{l} d^{n}l {\bar{f}}_{\Gamma}
       (\{l\}, \ldots; \{\sigma\})
\end{eqnarray}
where $\bar{f}_{\Gamma}(\{l\}, \ldots; \{\sigma\})$ denotes the 
integrand obtained from the underlying theory corresponding to the
graph $\Gamma$ and the dots refer to the LE parameters like 
external momenta, mass parameters and coupling constants. 
Other symbols are self-evident.

If the graph is totally convergent, then $\Gamma^0$ contains no 
UV ambiguity and the limit operation can be moved across all the 
internal integrations to act upon the integrand to give the product 
made from the propagators and vertices given by the present 
QFTs. But if there is any potential UV ill-definedness with any 
internal integration, one can no longer push the limit operation 
across this integration. Then following our treatment for the 
one-loop case, suppose that a graph $\Gamma$ contains at 
least an overall divergence, we proceed like the following, (we 
will use in the following $\omega_{\gamma}-1$ to denote the 
overall divergence index \cite {Shirkov} for any graph $\gamma$ 
and $\{l\}$ to represent the internal momenta and all the partial 
differentiation operators and their 'inverse' (which will be in the
following denoted by $\partial_{\omega_\gamma}^{-1}$) act 
upon the momenta only external to the very internal integration 
of the graph under consideration)
\begin{eqnarray}
&  & \Gamma^0(\ldots;\{c^0_{i}\})={\bf L}_{\{\sigma\}} \int
       \prod dl \bar{f}_{\Gamma} (\{l\},\ldots; \{\sigma\}) \nonumber \\
&\Rightarrow&\partial_{\omega_\Gamma}^{-1}{\bf L}_{\{\sigma\}} 
                    \int \prod dl \partial^{\omega_{\Gamma}} 
                    \bar{f}_{\Gamma} (\{l\},\ldots; \{\sigma\})\nonumber \\
&=&\sum_{\{\gamma\}=\partial^{\omega_{\Gamma}}\Gamma}
     \partial_{\omega_\Gamma}^{-1} {\bf L}_{\{\sigma\}} \int 
    \prod dl \bar{f}_{\gamma} (\{l\}, \ldots; \{\sigma\}).
\end{eqnarray}
Here we note that the differentiation wrt the external parameters
'created' a sum of graphs $\{\gamma\}$ (without overall divergence) 
from the original graph $\Gamma$. (Note that any overall 
overlapping divergence is hence killed by the ${\partial}^{\omega}$ 
operation, only non-overlapping divergences remain, i.e., the 
overlapping divergences are disentangled \cite {CK}). If there
is no more ill-definedness (in any subgraph), one can move the 
limit operator across all the internal integrations to act directly 
upon the integrands $\bar{f}_{\gamma} (\{l\}, \ldots; \{\sigma\})$ 
just like the overally convergent graphs. Now one can carry out 
all the loop integrations without any trouble for each graph 
$\gamma$ and then sum them up and finally apply the 'inverse' 
operator wrt the parameters (usually momenta) external to the 
graph $\Gamma$ (and each $\gamma$). 

But if there are still ill-definedness with some subgraphs for each
$\gamma$, then we can not move the LE limit operator across
all the loop integrations. In this case, each graph in the set 
${\partial}^{\omega_{\Gamma}}\Gamma$ can be expressed 
as a 'product' of divergent (at least overally divergent) but 
disconnected subgraphs (each subgraph itself may contain 
overlapping divergences), the LE limit operator does 
not commute with the loop integrations associated with these 
subgraphs though the other parts complement to these subgraphs 
can be applied with the LE limit operator. That is, the LE limit 
operator crossed all the other parts and stopped before the 
divergent subgraphs. In formula, for each graph $\gamma$,it is
\begin{eqnarray}
& &\partial_{\omega_\Gamma}^{-1} \left \{ {\bf L}_{\{\sigma\}}
     \int \prod dl \bar{f}_{\gamma} (\{l\}, \ldots; \{\sigma\})
     \right \} \nonumber \\
&=&\partial_{\omega_\Gamma}^{-1} \left \{\int \prod d 
       \overline {l^{\prime}} g_{\gamma/{[\gamma^{\prime} ]}}
      (\{\overline{l^{\prime}}\}, \ldots)\right.  \nonumber \\
& &\left. \times {\bf L}_{\{\sigma\}} \left [ \prod_{\gamma^{\prime}_{j}}
      \int \prod_{i\epsilon \gamma^{\prime}_{j}} 
      dl^{\prime}_{i} {\bar {f}}_{\gamma: \gamma^{\prime}_{j}}
      (\{l^{\prime}\}_{\gamma^{\prime}_{j}},\ldots; 
      \{\sigma\})\right ] \right \} \nonumber \\
&=&\partial_{\omega_\Gamma}^{-1} \left \{ \int \prod d 
      \overline{l^{\prime}}g_{\gamma/{[\gamma^{\prime}]}}
      (\{\overline{l^{\prime}}\}, \ldots) \times  \right. \nonumber \\
& & \left. \prod_{\gamma^{\prime}_{j}} \left [ {\bf L}_{\{\sigma\}}
   {\overline{\Gamma}}_{\gamma: \gamma^{\prime}_{j}} 
    (\ldots;\{\sigma\}) \right ] \right \}, \\ 
& &(\bigcup_{\gamma^{\prime}_{j}} 
     \{l^{\prime}\}_{\gamma^{\prime}_{j}}) \bigcup
     \{ \overline{l^{\prime}}\}=\{l\}, \\
& & [\gamma^{\prime}] \bigcup \gamma/{[\gamma^{\prime}]} 
      = \gamma, \ \ \ [\gamma^{\prime}]= \prod_{j} \gamma^{\prime}_{j},
      \nonumber \\
& &\gamma^{\prime}_{j}\bigcap \gamma^{\prime}_{k} =0, \ \
     for  j \neq k,
\end{eqnarray}
where all the dots in the expressions refer to the parameters 
'external' to the loop integrations for the subgraphs (i.e., to the 
$\gamma^{\prime}_{j}$'s)--they are the external parameters for 
the original graph $\Gamma$ (also for all the graphs in 
$\{\gamma\}$) {\it and} the internal momenta in the set
$\{\overline{l^{\prime}}\}$. 
$\overline{\Gamma}_{\gamma: \gamma^{\prime}_{j}}$ 
refers to amplitude derived from the underlying theory that 
corresponds to each subgraph $\gamma^{\prime}_{j}$ contained 
in $\gamma$. {\it Since some loop momenta are 'external' to 
certain subgraphs, one can not first carry out these loop 
integrations before the ill-defined subgraphs are treated and 
the loop integrations for these subgraphs are done. } This is 
in sheer contrast to the totally convergent graphs where the 
loop integration order does not matter. 
 
As the ill-defined subgraphs in $[\gamma^{\prime}]$ are 
disconnected with each other, we now treat each of them 
separately as a new 
'total' graph just like what we have done with the total graph 
$\Gamma$ starting from Eq.(11). Then we go through the 
procedures from Eq.(11) to Eq.(14) till we meet with new 
disconnected and ill-defined subgraphs that are in turn to be 
treated as before. Finally, we will go to the smallest 
subgraphs that are completely convergent. 
Now we can finally move the LE limit operator across all the 
loop integrations to get the integrands totally expressed with
propagators and vertices given by the effective theories and we
can begin to perform all the loop integrations in such an order 
(a 'natural' order from our treatment): First, perform the loop 
integrations for these smallest convergent subgraphs, then by 
construction perform the 'inverse' differentiation operator wrt the
momenta (or masses,etc., depending on technical convenience) 
external to these smallest subgraphs, and we will obtain 
ambiguous but finite expressions in terms of these 'external' 
parameters as generalized vertices for the higher level subgraphs 
which again by construction are convergent ones even with the 
generalized vertices. Secondly, go backward to carry out the 
loop integrations for these next-to-smallest subgraphs first and 
then perform the 'inverse' operation if any associated with these 
subgraphs, we will again arrive at generalized vertices for still 
'larger' subgraphs with more ambiguities appearing with the 
'inverse' operation. [It is worthwhile to note that at each level of 
the subgraphs, the loop integrations are guaranteed to be 
convergent due to Weinberg's theorem \cite {weinth}]. The 
process goes on till all loop integrations and all 'inverse' 
operations are done.

The resulting expression will be a definite nonlocal functions plus 
nonlocal ambiguities (due to subgraph ill-definedness) and local 
ambiguities if $\Gamma$ is suffering from overall divergence,
\begin{eqnarray}
& &\Gamma^0(\ldots; \{c^0\}) \Rightarrow
      \Gamma(\ldots;\{C\})   \nonumber \\
&=& \Gamma^{npl}_{0}(\ldots)+
     \Gamma^{npl}_{1}(\ldots;\{C^{\prime}\})+
     N^{\omega_{\Gamma}}(\ldots;\{\bar{C^{\prime}}\}), \\
& &\{C^{\prime}\} \bigcup \{\bar{C^{\prime}}\}=\{C\}.
\end{eqnarray}
Here again we used $ N^{\omega_{\Gamma}}$ to denote the 
polynomial containing the ambiguities ($\{\bar{C^{\prime}}\}$) 
appearing due to the overall divergence. Others are nonlocal 
functions. Different from the single loop case, there are 
nonlocal ambiguities in this multiloop graph suffering from 
subgraph divergences (as evident from our treatment) in 
addition to the nonlocal definite part and the local ambiguous 
part. The result we obtained ($\Gamma (\ldots;\{C\})$) is not 
what we are really after ($\Gamma^0(\ldots;\{c^0\})$), but that 
is the best we can do with the present QFT. One can easily 
see that this formulation will yield the expressions shared 
with BPHZ construction \cite {Shirkov,BP} if one have correctly 
constructed it with all the loop integrations done. Moreover, 
like any conventional scheme, the BPHZ expressions should 
correspond to certain choices of the constants $\{C\}$ for 
$\Gamma(\ldots;\{C\})$ in our approach. That is, we can 
provide a universal formulation for all the Reg and Ren 
schemes at least in the perturbative framework.

It is important to point out again that the amplitudes must have 
been parametrized by the constants $\{c^0\}$ representing the 
influences of the underlying theory (as we already addressed) in
addition to the LE phenomenological parameters (the dots that
represent collectively the external momenta, masses and 
couplings). Hence, it is not difficult to see that the important 
roles of these constants could not be guaranteed by simply 
replacing them with {\it one} running scale as done in the 
usual Ren schemes. 
Thus, the real problem is how to find the 'truth'--the constants 
$\{c^0\}$--or how to define the $\{C\}$'s as we stressed in 
section II.

Here some remarks are in order.

{\bf A}. It is evident that overlapping divergences are just 
automatically resolved in our approach, there is nothing special
about it. This is because the differentiation operators just 
'kill' the overlapping ill-definedness by 'inserting' internal lines
and vertices to reduce the overall divergence. Thus one need not
worry about them any more. This is the utility derived from the
differentiation wrt external parameters (momenta, masses or
other massive parameters that might appear in the LE 
propagators) \cite {CK}. This also dispenses the laborious
construction of the counter terms when there are overlapping
divergences in the usual Ren framework.

{\bf B}. Since the amplitudes constructed from the underlying 
theory are definite (even their LE corresponding graphs are 
ill-defined), they must remain unchanged under any linear 
transformations of the internal integration variables provided that 
the determinants of the jacobians for the transformations are 
identities. In our treatment of the ill-defined graphs, since every 
loop integration actually performed is convergent,  
these transformations do not alter the results of the loop 
integrations. Due to the 'inverse' operator, these linear 
transformations of the integration variables will at most change
the ambiguous constants. But that does not matter at all, 
since these constants are yet to be determined, they can well 
absorb any finite changes. Or, it merely leads to Reg effects.
This observation implies that one should not worry about the 
variable shifting and routing of the external momenta that 
belong to the transformations just described.

{\bf B1}. An immediate corollary to this observation is that, the 
chiral anomaly, which is conventionally interpreted as due to 
the variable shifting in relevant linearly divergent amplitude, 
must have been due to other definite properties. Otherwise, 
if it were totally due to the local 
ambiguities, one can well remove them away by choosing 
appropriate definitions of the constants (or appropriate Ren 
conditions). Our direct calculation shows that \cite {JF,JF1}, 
one kind of definite rational terms (independent of masses) 
originated the chiral anomaly. Since they are nonlocal and 
unambiguous, one can not attribute them simply as UV effects 
and remove them. The trace anomaly is also shown to be 
originated by such kind of rational terms \cite {JF,JF2}. To our 
best knowledge, this nontrivial structure (independent of the 
UV ambiguities) has never been noted before in the old Ren 
framework.

{\bf B2}. Another utility derived from the observation is that, 
one can choose routings of the external momenta to be as 
simple as possible to make the treatments of  an ill-defined 
multi-loop amplitude as easy as possible. For the single loop 
cases, sometimes one may only focus on the parts of the 
amplitude that are really divergent. This may yield fewer 
ambiguities.

{\bf C}. As we have seen that our treatment can lead to a
universal 'parametrization' of the Reg effects ( at least in
perturbative approach), one can, in actual calculations,
employ a specific Reg that saves labor of the calculations,
and replace all the regularization parameter dependent
parts in the resulting expressions with general ambiguous
polynomials with correct order of power.  But the definitions
of the ambiguities should be done following our treatment.

{\bf D}. Although we do not need the detailed knowledge
about the underlying theory in our approach for the QFTs, 
it is now very clear that, the present formulation of QFTs
is incomplete to define everything for relevant phenomena.
Thus one has to supplement in the usual approaches
something by hand--the introduction of a Reg. 
Conceptually, they are necessarily 
artificial substitutes for the underlying structures.

By now, it is clear that all the LE effective theories (any model 
in any dimensional space-time) can be treated in this way, and 
they should be UV finite . The only and yet difficult problem is 
that we are facing ambiguities and we do not know the true and 
final answer about them. Generally, as we have discussed in 
the section II, we may first impose some novel symmetries and 
invariances on the amplitudes to reduce the ambiguities to 
certain degree, then one has to resort to the experimental 
physics data. In principle, if this idea is correct, we can
'reproduce the truth' due to the structural relations between 
the Feynman graphs as follows.

For convenience we divide all the graphs(or FAs ) into three 
classes: (A) overall-divergent ones; (B) overall-convergent ones 
containing ill-defined subgraphs; and (C) the rest, totally well 
defined graphs. We need to resolve all kinds of ambiguities in 
classes (A) and (B). First let us look at class (B). For a graph 
in this class, one would encounter nonlocal ambiguities due to 
the subgragh ill-definedness. Such graphs must correspond to 
certain physical processes as they carry more external lines, 
thus, the ambiguities in their nonlocal expressions will in 
principle be fixed or removed by relevant experimental data, 
that is, {\sl the ambiguities in the subgraphs are also 
constrained by "other graphs". } So, with the experimental 
data, the nonlocal ambiguities (from the local ambiguities of the 
subgraphs in fact) are in principle completely fixed or removed. 

To solve the problem with class (A), we note that class (A) can 
all be mapped into class (B) as subgraphs of the latter,  then 
the resolution of the ambiguities in class (A) follows immediately. 
Thus, to our surprise, due to the Feynman graph structures of 
the whole theory all the potential ambiguities or divergence's 
should not materialize at all if the theory does not suffer from 
structural inconsistency and one can in principle use the 
experimental data sufficiently. The important thing is that this 
resolution (in principle) is only valid for the complete theory, 
that is, a nonperturbative 
conclusion rather than a perturbative one. 

Then, the problem becomes: can these "definitions" be 
consistently done? The answer will certainly depend on model 
structures, then a new classification for the QFT models for 
certain energy ranges based on such consistency shows up: 
category one ( $FT_{I}$ here after) with consistent "definitions" 
implementable, category two ($FT_{II}$) without such consistency. 
In a sense, category two appears due to our incomplete 
'assembling' of the Green functions. Of course $FT_{I}$ interests 
us most, but as the energy range of concern extends upward, the 
set $FT_{I}$ will "shrink" while the set $FT_{II}$ will swell. The 
final outcome of this "move", if accessible at all, should be the 
final underlying theory unique up to equivalence (like the present 
situation in superstring theories \cite {JH} somehow). As for the 
relation between this classification and that judged by 
renormalizability, we can claim nothing rigorously before further
investigations is done. Intuitively QED, etc. seem to belong to
category one $FT_{I}$.

It is time to discuss a formulation based on Wilson's picture 
\cite {pol}. We note that Wilson's picture is basically the same 
as the one we used as a postulate. But it is crucial to note that 
the formulation of Ref \cite {pol} is based on such an 
interpretation of the Wilsonian picture, i.e., the content of the 
low energy physics is independent of the short distance theory 
up to parameter redefinition effects, which in turn leads to a 
formulation that uses this interpretation (equivalent to imposing 
the Ren group (RG) invariance) as technical starting point. 
However, from our 
discussions above, this is an {\it ad hoc } assumption as the 
Ren conditions affect physics and the independence of the low 
energy theories upon the short-distance theory scale (acting as 
a cutoff) does not necessarily mean that the effective theories 
are independent of the Ren conditions. In our point of view, the 
procedure of fixing the ambiguities or the Ren conditions is the 
most important thing as {\it different choice leads to different 
physics ! } (We will discuss some evidences from quantum 
mechanics in next section). It is also not justified to use just one 
simple energy scale to parametrize all the short distance theory's 
influences on the LE theories let alone to simply cut off the 
{\it 'HE' modes given by the LE models.}

As we have pointed out, the Feynman Amplitudes or the 1PI
functions are generally parametrized by more than one constants 
(I will refer to them as 'radiative constants (RCs)') in addition to 
the phenomenological ones ( classical masses and couplings). 
If the changes in the RCs could be completely compensated by 
that in the phenomenological ones (which is only possible for 
rather special kind of models ), then we might implement a 
redefinition invariance of the constants for the FAs like in the 
RG case. Such an 'invariance' needs verification rather than
being simply imposed upon and could not be simply
parametrized by one parameter like in RG equation (RGE). 
Otherwise, beyond the utility as a technique for partial 
summing to go beyond the perturbation, RGE, if exists, 
should only correspond to real physical symmetry. For the 
critical phenomena, the symmetry is that of the scaling law, 
and there are physically meaningful infinities--the infinite 
correlation lengths. For the general LE particle physics far 
from the thresholds where new physical excitations or modes 
may appear, there is no sensible scaling symmetries nor 
sensible physical infinities. Hence, no sensible symmetry can 
lead to RGE in such general cases where most QFTs work well.
For the deep inelastic scattering (DIS), where thresholds for 
new physics come closer, the dynamics is about to undergo
a 'phase transition', at least approximate scaling symmetry may 
show up and RG like equation tends to be real thing. Such 
results can in principle be achieved in any approach if one has 
correctly performed the treatment of the ambiguities. In fact, the 
DIS is just the arena where RGE built up its reputation in high 
energy physics. 

We wish to note that since there is no room for divergence and
hence no room for bare parameters in our approach, the so-called
mass scale hierarchy problem in the Standard Model \cite {Hie}
may become less serious if one adopts our proposal, the 
'fine-tuning' problem will be superceded by the determination
of the radiatively-arised constants (which should be taken as
the influences from the underlying theory, just what we discussed
above)from physical requirements. Thus, the original need for the 
supersymmetry \cite {SUSY} to cancell certain UV divergences
in order to overcome this 'fine-tuning' \cite {Hie} should be 
reexamined (within our new approach) in its physical relevance 
rather than in infinity removing uses.

\section{Applications to Nonperturbative Examples}

From the presentation above, it is clear that our approach works 
in principle for any model, whether it is a QFT or not. The key
observation that the UV ill-definedness is caused by illegitimate
order of 'operations' is valid for both perturbative framework and
the non-perturbative ones, see Eq.(1). That is to say, our 
approach should apply to nonperturbative calculations, with 
perhaps some technical modifications. 

It is important in the usual Ren methods that one has enough 
classical parameters to absorb the divergences for problems 
in study. Once there are more divergent integrals to be
compensated, one can by no way remove all the divergences 
and the conclusions thus obtained were in fact questionable, 
especially in the nonpertubative contexts. In our point
of view, it implies the rationale and techniques of the old Ren
schemes are simply bad 'substitutes'. The results obtained 
in these schemes might be incorrect or even irrelevant to the 
physical phenomenon under concern. In other words, 
nonperturbative problems can be critical touchstones for 
these schemes. Of course, they also provide tests for the 
correctness and reasonableness of our approach. That is 
why I would like to discuss these examples.

Recently, the cutoff Reg and Dimensional Reg are compared 
in nonperturbative context in quantum mechanics with
Delta-potential problems \cite {QMDR}. Quantum mechanics 
with Delta-otentials can be natural framework for contact 
interactions in nuclear physics, molecular physics and 
solid-state physics. We are especially interested in the 
case in nuclear physics, as it is recently a hot topic initiated 
by Weinberg's suggestion \cite {WeinEFT} that the technology 
of effective field theory (EFT \cite {EFFT}, not exactly that 
proposed above) could be used to describe LE nuclear physics 
phenomena. The resulting 
theory is a non-relativistic quantum mechanics with Delta-
potentials, which are in fact singular in the short-distance 
behavior by birth. According to our discussions above, we 
should first bear in mind that, when there is problem of
unphysical UV infinities, it means that the LE effective models 
must have failed in the higher energy end. Or the theory is 
ill-defined and it is illegitimate to simply work with the 
propagators and vertices ( or Green functions and potentials) 
given by the very LE models, nor should one introduce any 
{\it ad hoc } Regs without taking care of its unphysical 
ingredients. Great care must be taken wrt the Reg effects. 
Thus the inequivalence between the cutoff Reg and 
Dimensional Reg exhibited in Ref. \cite {QMDR} well 
evidenced the correctness of our arguments. Since no Reg 
and Ren scheme is superior to the others as long as UV 
infinities may appear or even irremovable within the very 
scheme, the conclusions arrived at with such Reg and Ren 
schemes need reexamination. 

Now let us show how to treat the problem within our approach. 
Generally, the Lippmann-Schwinger equation for $T$-Matrix in 
the simple two-body problems formally reads (we follow the 
notation conventions of Ref. \cite {QMDR})
\begin{equation}
T(p^{\prime},p; E)=V(p^{\prime},p)+ \int 
\displaystyle\frac{d^d k}{(2\pi)^d} V(p^{\prime},k ) 
\displaystyle\frac{1}{E^{+}-k^2/(2\mu)} T(k,p; E), 
\end{equation}
where $E^{+}$ is $E + i \epsilon$, with $E$ non-negative, and 
$\mu$ denotes the reduced mass in the two-body problem. In 
our point of view, this equation is not well-defined and should 
be written as the LE limit of that derived from the more 
fundamental underlying theory which is unavailable to us by 
now. [ We should note that, the underlying parameters will be 
always denoted as $\{\sigma\}$. For different problems or 
different LE ranges, the contents may differ, this is due to that 
for lower and lower energy level, some
modes which are themselves LE modes for still higher energy
ranges become relatively 'high energy' ones and inactive in the
more lower energy ranges, and the phenomenological 
parameters for these 'HE' modes become ( for the more lower 
energy dynamics) 'underlying' ones.] So in our language, 
Eq.(17) should be 'corrected' as
\begin{eqnarray}
T(p^{\prime},p; E ; \{\sigma\}) &=& V(p^{\prime},p ; \{\sigma\}) +
\int \displaystyle\frac{d^d k}{(2\pi)^2} V(p^{\prime},k ; \{\sigma\})
\nonumber \\
& & \times G(E^{+}-k^2/(2\mu); \{\sigma\}) T(k,p;E; \{\sigma\}), \\
V(p^{\prime},p)&\equiv&
{\bf L}_{\{\sigma\}} V(p^{\prime},p ; \{\sigma\}), \nonumber \\
\displaystyle\frac{1}{E^{+}-k^2/(2\mu)}&\equiv&
{\bf L}_{\{\sigma\}} G(E^{+}-k^2/(2\mu); \{\sigma\}).
\end{eqnarray}
Eq.(18) is now well-defined in the underlying theory.Thus 
Eq.(17) is correct only when there is no UV infinities (again 
as before we assume no IR problem is in concern as is 
indeed the case in the following discussions for the 
Delta-potential problem) so that the LE limit operator can 
cross the internal momentum integration (summation over 
intermediate states) and act on everything. Otherwise we 
have to find a legitimate way to let the LE limit operator cross 
everything (acting on everything) so that we can calculate with 
the objects given by the LE theories.

In the case of Delta-potential, $V(p^{\prime},p)=C$, but the 
$V(\ldots; \{\sigma\})$ is generally a nonlocal potential before 
the LE limit is taken. To be rigorous,  we write formally
\begin{eqnarray}
T(p^{\prime},p; E ;\{c^0\})&=& C+
{\bf L}_{\{\sigma\}} \left \{ \int \displaystyle\frac{d^d k}{(2\pi)^d} 
V(p^{\prime},k;\{\sigma\})\right. \nonumber \\
& &\left. \times G(E^{+}-k^2/(2\mu); \{\sigma\})
T(p^{\prime},p; E ;\{\sigma\}) \right \},
\end{eqnarray}
and it is not generally legitimate move the 
$V(\ldots;\{\sigma\})$ out of the integration to be directly subject 
to the LE limit operator--which is exactly what was done in the 
conventional calculation (with only the propagator regularized)--
and it is definitely illegitimate to apply the LE limit operator to 
all the other objects before the integration is done. Thus, in 
principle, even when the LE potential is local (of course 
$V(\ldots;\{\sigma\})$ is nonlocal), 
it might be dangerous to simply reduce Eq.(18) to an algebraic 
one. Only when the ill-definedness is mainly caused by
$1/(E^{+}-k^2/(2\mu))$ (i.e., it differs greatly from
$G(\ldots;\{\sigma\})$ in the UV region where $V(\ldots)$ differs
less from $V(\ldots;\{\sigma\})$, we could pull out the true 
potential to subject it directly to the action of the LE limit 
operator. In other words, to put Eq.(18) (a correct formulation for 
Eq.(17)) or Eq. (20) into an algebraic one requires quite 
nontrivial properties of the potential and the propagator, 
which {\it the usual analysis failed to note. } To focus on the 
main point, we temporarily assume this condition is satisfied, 
then we have the well-defined form of the algebraic equation for 
the $T-$matrix (which is now parametrized by the new constants 
$\{c^0\}$ from the LE limit in addition to $E$),
\begin{equation}
\displaystyle\frac{1}{T^{on}(E;\{c^0\})} = \displaystyle\frac{1}{C}
- I(E;\{c^0\}),
\end{equation}
with 
\begin{equation}
I(E;\{c^0\})={\bf L}{\{\sigma\}} \int \displaystyle\frac{d^d k}{(2\pi)^d}
G(E^{+}-k^2/(2\mu);\{\sigma\}).
\end{equation}
Now we can employ the technique described in sections II 
and III to calculate the integrals, i.e., first differentiate 
$G(E^{+}-\ldots; \ldots)$ wrt $E^{+}$ (which is the 'external' 
parameter in the integral) for appropriate times, secondly 
perform the LE limit legitimately and carry out the integral thus 
obtained, finally do the 'inverse' operation wrt $E$ and we find 
the followings (note that here one differentiation wrt $E$ 
reduces the divergence degree by two)
\begin{eqnarray}
&  & I_{d;odd} (E;\{c^0\}) \Rightarrow I_{d;o}(E;
     \{c^{\prime}\}) \nonumber \\
&= & -i\displaystyle\frac{2\mu {\pi}^{d/2+1}}{\Gamma(d/2) (2\pi)^d} 
       (2\mu E)^{d/2-1} + N^{[(d-1)/2]} (2\mu E;\{c^{\prime}\}); \\
&  & I_{d;even} (E;\{c^0\})\Rightarrow I_{d;e}(E;
     \{c^{\prime}\})\nonumber \\
&= & \displaystyle\frac{2\mu \pi^{d/2}}{\Gamma(d/2) (2\pi)^d} 
       (2\mu E)^{d/2-1} \ln (2\mu E/c^{\prime}_0) + 
        N^{[d/2]} (2\mu E;\{c^{\prime}\})
\end{eqnarray}
with $\{c^{\prime}\}$ being arbitrary constants--the ambiguities.
These expressions can again be viewed as universal 
parametrizations and compared with that given in cutoff Reg
and dimensional Reg schemes (C.f. Ref \cite {QMDR}) with the 
latter ones as special cases. 

In terms of the ambiguous (but {\it finite }) integrals given by 
Eq.(23,24), the $T-$matrix is now parametrized by 
$\{c^{\prime}\}$ in addition to $E$ like
\begin{equation}
\displaystyle\frac{1}{T^{on}(E;\{c^{\prime}\})} = 
\displaystyle\frac{1}{C} - I_{d;\ldots}(E;\{c^{\prime}\}).
\end{equation}
Again we need to fix the constants $\{c^{\prime}\}$ rather than 
to renormalize the interaction constant $C$.

It is easy to see that following the normalization condition of 
Ref. \cite {QMDR}, we can well reproduce the result derived
by Weinberg \cite {WeinEFT} in two or three dimensional 
space-time. However, there seems to be no necessary 
constraints on the phenomenological constant $C$ as it is 
physical rather than 'bare' in our approach. Thus, to us good, 
this LE framework \'a la Weinberg \cite {WeinEFT} serves equally 
well for both the attractive interactions and the repulsive ones if 
one adopts our approach, contrary to the conclusions that EFT 
framework failed in the repulsive cases where the LE models are 
believed to be trivial \cite {Beg,Cohen,PC}. As a matter of fact, 
the triviality, if investigated in our approach, can not be the 
unique conclusion provided one carefully fix the ambiguities. 
The nontriviality of the Delta-potential dynamics has also been 
investigated by Jackiw \cite {Jackiw}. In our point of view, the 
triviality conclusion is flawed as the Reg factors had not been 
seriously taken into account (which is common in many 
conventional studies, and it is hard to be avoided if one does 
not adopt the viewpoints and technical approach proposed 
above), especially when a Reg makes the results containing 
irremovable infinities (which is just 
the case for the cutoff Reg scheme adopted in Ref. 
\cite {QMDR,Cohen,PC}) as we remarked above.

This problem can be attacked from another angle. Jackiw
had already pointed out \cite {Jackiw} that the Hamiltonians 
for such models are not automatically Hermitean but need 
self-adjoint extension. This has already been dealt with by
mathematicians in the operator theory \cite {AG} and has also 
been extensively discussed by physicists \cite {Albeverio} in a 
number of approaches such as boundary value conditions at 
the short-distance limit, Dirichlet quadratic form approach, 
nonstandard analysis method, resolvent method and others 
(please refer to \cite {Albeverio} for a comprehensive list of the 
literatures for these approaches). Each approach, if viewed from
our standpoint, amounts to a way of trying to retrieve the lost 
information about the UV underlying structures. The key point 
is, in such cases, the self-adjointness of the Hamiltonian is
never self evident and beyond examination. That is, in contrast 
to the normal case, the contact potential problem \'a la
Schr\"{o}dinger equation {\it is ill-defined. } The resolution of 
the problem gives rise to a family of self-adjoint extensions of 
the original Hamiltonian operator parametrized by an additional 
constant, which upon different choices leads to different or 
inequivalent (LE) physics \cite {QM}. This additional 'family' 
parameter is just the constant that will surely be predicted from 
the LE limit operation in our approach, corresponding to the 
ambiguity whose definition requires most attention as stressed 
for times in section II and III. As a matter of fact, there is an 
approach that is technically quite similar to ours here, the one 
based on resolvent formalism \cite {Grossmann} where an 
important object is defined through an equation in which it 
appeared in a form differentiated wrt the 'external' 
parameter--the resolvent variable (energy). Thus this important 
object is only defined up to an additional parameter---the family 
parameter in other formalism--which is to be determined by 
other input, just like in our approach.

Now we can understand why the conventional Ren approaches 
failed in such problem. First, the Regs used ( often the cutoff 
Reg scheme) may in the first place introduces unphysical 
infinities that might spoil the results esp. in the nonperturbative 
case as demonstrated in Phillips {\it et al } 's work 
\cite {QMDR} and discussed above (they are not guaranteed to have
correctly parametrized the UV underlying dynamics as they are
simply artificial substitutes!). Secondly, the physical 
definition of the additional constant(s) would be impeded by the 
irremediable remaining UV infinities. Thirdly, the conclusions 
thus reached were either irrelevant to the LE physics or 
questionable. So, the proofs that the effective range of the 
Delta-potentials is non-positive based on the cutoff Reg 
\cite {Cohen,PC} are flawed ones. The proof should be done 
in several independent approaches and each should take good 
care of the definition of the 'free' parameter and its implications. 
Since, {\it the self-adjointness and hence the unitarity of such 
models can not be simply assumed and started with } and 
one should instead exert quite nontrivial efforts to define them 
carefully, we hold that the status about the utility of these LE 
contact interaction models remains unsettled just as Jackiw 
indicated in his work \cite {Jackiw}. 

To this stage, we could see that, the $\lambda \phi^4$ theory 
in 3+1-dimensional space-time, which is conventionally held 
as trivial, should now be understood as a LE effective model 
just like any other QFTs as QED, QCD,etc. The only question 
about its utility lies in whether it could be consistently defined 
wrt the ambiguities. To this end, we view it an unsettled question. 
The conclusions drawn from the lattice approach, in our eyes, 
are questionable as it is again a cutoff like Reg scheme that 
suffers from severe UV divergences that are irremovable (the 
lattice approach is non-perturbative).

Now it is evident that our approach is more reasonable than 
the conventional Ren frameworks both from physical rationality 
and from the capability of dealing with the nonperturbative 
ill-definedness quite simply as well as efficiently where the 
conventional approaches could hardly reach reasonable 
conclusions as they are often heavily plagued by the bad Reg 
schemes and hence by the irremediable UV infinities.

I would like to mention a recent investigation \cite {GEP} on
Higgs particles in nonperturbative context employing the 
approach we proposed here (see also \cite {YYY,JF}). The 
results thus obtained were neat and clear, comparing with 
that performed within the old Ren framework. Especially, the 
physical pictures are different from that using the old Ren, 
which is now easy to see from the discussions above.

Another work by Dai {\it et al } \cite {DAI} also implied that the 
conventional quantum mechanical framework failed in the case
of the presence of singular (UV) potential. Care must be taken
in the construction of complete and orthogonal eigenvectors 
with real eigenvalues where again a 'free' parameter arised 
naturally in the course of the construction \cite {Case}.

One may expect that great ease can be found in employing 
our approach or its equivalents (in any form known or 
unknown) in his/her studies in the nonperturbative contexts
and the outcome of this act would be quite different and
significant. Moreover, within our approach, those 
phenomenologically oriented models which are 
unrenormalizable in the usual Ren schemes could become
quite tame ones and be of great help. One can test it with the 
NJL model and chiral perturbation theory \cite {CPT} and even 
with gravity \cite {Dono}. We also note that the principally 
nonperturbative effective action formalism \cite {Cornwall} 
which is widely used, once equipped with our strategy, will 
greatly help to illuminate the topics concerned and to produce 
quite different but nonetheless physical conclusions which 
are often unattainable within the old schemes.
 
\section{About IR Problem}

Now let us consider the infrared (IR) problems. Conventionally
we have the Kinoshita-Poggio-Quinn theorem 
\cite {Kino,PQ,Muta} and the Kinoshita-Lee-Nauenberg 
theorem \cite {Kino,Muta,LN} to take care of them in off-shell 
Green functions and on-shell Green 
functions (or S-matrix) respectively for QCD and the like. As 
they are obtained with the assumption that the UV ambiguities 
(or divergences) have been removed, we may expect the 
same hold for $FT_{I}$ in our treatment where the finite yet 
ambiguous constants take place of the divergences and the 
subsequent subtractions, at least for the gauge theories in this 
class. The IR problem for gauge theories is in fact due to the 
degeneracy of charge particle states "wearing"  soft boson 
clouds \cite {Kino,Muta,LN}and its deeper origin is shown to 
be the conflict between gauge symmetry and Lorentz 
invariance \cite {Haag}. Hence the IR issue would contribute 
something nontrivial to the physical requirements for the set 
$FT_{I}$. In fact, the IR problem for QCD concerns directly 
dynamical color confinement \cite {Confine} which is in turn 
closely related to dynamical chiral symmetry breaking and 
finally to the complicated nonperturbative vacuum structure, 
these all can, once resolved, lead to quite significant structural 
constraints on the RCs.

Here, I would like to suggest another way of thinking which 
aims at again a general framework for dealing with unphysical 
IR infinities with more physical rationale. Due to our discussions 
in the first section, we have seen that the existence of the 
fundamental underlying theory implies that the QFTs we have in 
hand are simplified LE limit, hence the spectra structures given 
by these effective models deviate from the ones given by the 
underlying theory, i.e., the UV ends of the effective spectra are 
incorrect, and the effective modes break up or new fundamental 
modes become active in UV ends. The similar reasoning works 
for the IR divergence. For a complete representation of the 
world, we should expect that the underlying
theory is also well defined in the IR sector. (Our discussions in 
the introduction about the spectra are partial as we deliberately 
omitted the IR issues to focus on the UV structures.) It is
conceivable that the phenomenological LE models give wrong 
information of the IR end spectra signaled by the unphysical IR 
infinities.

The underlying theory, 'postulated' here, if exists, should 
contain all the nontrivial UV and IR structural information that 
each effective theory--at least ill-defined at one end, UV or 
IR--lacks and 'misses'. Then an interesting scenario dawns 
upon us: for each effective model dominating certain energy 
range (say, theory $I_{mid}$), there should exist two other 
effective models (or sectors) that are most adjacent to this 
model from the IR end and UV end respectively (say, $I_{IR}$ 
and $I_{UV}$). Then it is imaginable that the phenomenological 
parameters in $I_{IR}$ and/or $I_{UV}$ would at least quite 
nontrivially improve the status of the IR and/or UV behaviors of 
the theory $I_{mid}$. While on the other hand, the $I_{mid}$ 
contains what $I_{IR}$ (resp. $I_{UV}$) needs to improve its 
UV (resp. IR) behaviors. Put it another way, the active and 
'elementary' modes or fields in $I_{IR}$ will break up in 
$I_{mid}$ and give way to the new 'elementary' modes active 
in $I_{mid}$. Similarly, the 'elementary' modes in $I_{mid}$ 
will go 'hibernating' as the energy goes down while 'new' 
elementary modes 'emerge' to dominate spectra in $I_{IR}$. 
The relation between the elementary modes in $I_{mid}$ and
$I_{UV}$ is in principle just like that between those in $I_{IR}$
and $I_{mid}$. Of course, there may be modes active in several 
successive effective models, some may even be active and 
stable through all energy levels---the 'fossil' modes or fields we 
mentioned in the introduction. Evidently, the information 
about those 'elementary' modes in $I_{IR}$ and $I_{UV}$ 
missing from $I_{mid}$ (i.e., missing from the effective 
spectrum given by $I_{mid}$) can contribute to improve the 
IR and UV behavior of the latter. 

Thus, in a sense, both the IR modes and the UV modes and 
hence the associated phenomenological constants 
characterizing them 'underlie' a QFT (or more generally, a 
quantum theory) if this QFT is ill-defined in the IR and UV 
ends. In this report, we have shown a simple but powerful way 
of extracting finite results in spite of not knowing the true UV 
structures. To be complete, we should also work out a way 
to get rid of the IR unphysical infinities in the same spirit (but 
not necessarily with the same technique). The author does 
not have a solid idea for the answer right now. Through the 
discussions just made above, we can see that the 'underlying' 
structures (IR and UV) and the effective structures are in fact 
unified in an 'organic' way, they depend upon each other and
they contribute to each other. We might expect that more
efficient treatment of the IR troubles of a QFT would involve
the whole spectra properties (including the 'effective' range
and the UV end) in a quite sophisticated way. The whole 
'organic' as well as unified theory, if accessible, describes 
everything without any kind of ambiguity. 

Now, the main difficulty lies in the way to parametrize the 
effects from the underlying' IR sector upon the present 
theory's structures (through propagators, vertices or other 
components of the theory?) so that the IR ill-definedness
is signaled by the ambiguities. The effectiveness of the 
present QFT in the IR respect would be indicated by a limit 
operator saying that certain underlying parameters 
characterizing the IR end are 'vanishingly small'. Recently,
the use of duality and holomorphy in supersymmetric
field theories has led to tremendous progresses in 
obtaining nonperturbative results that are more well-defined
in the IR as well as the UV respect \cite {SW}. We hope
to be able to integrate this achievement as well as that of
the infraparticle \cite {Haag,Buch} into our future 
investigations on the solution of the IR ill-definedness 
\cite {JY1}.

\section{Discussions and Summary}

First, we note that our approach is rather general in concept 
and can be applied to any model (whether a field theoretical 
one or not) in any space-time ( whether Euclidean or 
Minkowskian). The true building blocks we work with are the 
various Green functions (parametrized by $\{\sigma\}$ in 
addition to the usual parameters). Thus, we may even work 
in a 'partial' model subject to future completion. Second, we 
also parametrized the 'elementary' fields of the effective 
theories to indicate that there are structures underlying these 
fields. This is also exhibited in Eq.(1) in the measures for the 
path integrals. Thus, it is also capable of dealing with the 
ill-definedness in the composite operators constructed from 
the field operators and that in the jacobians associated with 
the measures \cite {Measure}. The resulting expressions 
should be again given in terms of the phenomenological 
variables (constants and fields) and the finite constants 
(RCs, see section III) \cite {JY2}. Third, it is immediate to find, 
with the preparations above, that the Hilbert space associated 
with a LE theory should also be reconsidered with respect to 
the influences of the underlying parameters. This possibility is 
never discussed in the usual Reg and Ren framework (as they 
are only intermediate stage treatments) to the author's best  
knowledge. In our approach, this issue arises naturally. We 
should at least consider the implications of the underlying 
theory for the Hilbert spaces for the effective theories. Fourth, 
it would be interesting to integrate our approach with the BV 
anti-field formalism of quantization \cite {BV} that has been 
used to deal with the unrenormalizable theories quite 
recently \cite {gom}. 

We want to point out that, the conventional quantization 
procedure of fields is now subject to question. The 'elementary' 
commutator for a field (fermionic or bosonic) and its conjugate, if 
calculated (or formulated) from the underlying theory, must have 
been at least a nonlocal function(al) parametrized by the 
underlying parameters of the underlying theory and must have 
been closely related with the gravitational interaction and 
perhaps new fundamental ones, rather than a highly abstract 
Dirac delta function containing least information. In a sense, 
the incompleteness of the present QFTs or their ill-definedness 
is inherent in the present quantization procedure whose most 
elementary technical building block is the Dirac delta function 
(called as distribution by mathematicians) that is {\it extremely 
singular and can not be defined in the usual sense of function. 
} That the distribution theory works necessarily with test 
function space or appropriate measure, if viewed from physical 
angle, is equivalent to that we need more 'fundamental 
structures' in order for some singular functions to make sense, 
i.e., a necessity of introducing underlying theory or its artificial 
substitute--regularization. The constructive field theory 
approach, in this sense, also works with a regularization 
effected through the differential properties($C^{k}$) 
of the test functions. 

Thus, we have opened up many important topics that
deserve further serious investigations.

Our investigation here, significantly benefited from taking the 
QFTs as effective theories of a more fundamental underlying 
theory, suggests that the 'final answer', may not be a field 
theory \cite {Jac}, at least there is no hope for the field 
theoretical formulations like what we presently have. We 
temporarily refrain from making further remarks about the
relation between our approach and string theories.
We should also stress that, our approach does not resort to 
symmetries and invariances to get rid of the infinities, while 
the old Ren frameworks has to use them and thus are greatly 
limited \cite {Jac}. That is to say, we can use the models 
exhibiting less symmetric regularity for certain
phenomena without worrying about infinities any more.

In summary, we discussed in some detail the approach
recently proposed by the author and the important consequences
following from it. We have overcome many typical difficulties and
shortcomings associated with old Reg and Ren frameworks. The
method is simple and powerful in many respects.

\section*{Acknowledgement}
  I am grateful to Prof. S. Ying for his helpful discussions over a number of 
topics and to Prof. G.-j. Ni for his helps and encouragements in many
respects.


\begin{thebibliography}{99}
\bibitem {Jac} R. Jackiw, Report No. hep-th/9709212.
\bibitem {DR}  e.g., G. 't Hooft and M. J. G. Veltman, Nucl. Phys. 
{\bf B 44}, 189 (1972); L. Culumovic {\sl et al  }, Phys. Rev. 
{\bf D 41}, 514 (1990); D. Evens {\sl et al }, Phys. Rev. 
{\bf D 43}, 499 (1991); D. Z. Freedman {\sl et al }, Nucl. Phys. 
{\bf B 371}, 353 (1992); P. R. Mir-Kasimov, Phys. Lett. {\bf B 378}, 
181 (1996); J. J. Lodder, Physica {\bf A 120}, 1, 30 and 508 (1983);
H. Epstein and V. Glaser, Ann. Inst. Henri Poincare {\bf XIX}, 211 
(1973).
\bibitem {wein} S. Weinberg, {\it The Quantum Theory of Fields, }
Vol I, Ch. XII, Section 3, Cambridge University Press, Cambridge,
England,1995. The author is grateful to Professor J.Polchinski for this 
information.
\bibitem {YYY} Jifeng Yang, Report No. hep-th/9708104, 
hep-th/9801005; invited talk at the {\it XIth International Conference
(PQFT'98)} held in JINR, Dubna, Russia, July 13-17, 1998.
\bibitem {wil} K. G. Wilson, Phys. Rev D {\bf 4}, 3174, 3184 (1971);
K. G. Wilson and J. G. Kogut, Phys. Rep. {\bf 12}, 75(1974)
\bibitem {pol} J. Polchinski, Nucl. Phys. B {\bf 231}, 269(1984).
\bibitem {Bigi} I.I. Bigi, {\it et al, } Phys. Rev {\bf D 50}, 2234 
(1994).
\bibitem {Shirkov} N.N. Bogoliubov and D. V. Shirko v,{ \it Introduction
to the Theory of Quantized Fields, } 4th edition (Wiley, NY 1980).
\bibitem {JF}  Jifeng Yang, Ph.D. dissertation, Fudan University, 
unpublished, (1994).
\bibitem {CK} W. E. Caswell and A. D. Kennedy, Phys. Rev. {\bf D 25}, 
392 (1982).
\bibitem {LL} C. H. Llewellyn Smith, Phys. Lett. {\bf B 46}, 233 (1973).
\bibitem {QMDR} D. R. Phillips, S. R. Beane and T. D. Cohen, 
Report No. hep-th/9706070 and references therein.
\bibitem {QM}  P. Gerbert, Phys. Rev. {\bf D40}, 1346 (1989); 
M. Reed and B. Simon, {\it Methods of Modern Mathematical Physics, 
Vol II, } Ch 10 , p145 (Academic Press, New York, 1975). 
\bibitem {weinth} S. Weinberg, Phys. Rev. {\bf 118}, 838 (1960).
\bibitem {BP} N.N. Bogoliubov and O. S. Parasiuk, Acta. Math.
{\bf 97}, 227 (1957); K. Hepp, Com. Math. Phys. {\bf 2}, 301 (1966);
W. Zimmermann, Com. Math. Phys. {\bf 11}, 1 (1968).
\bibitem {JF1} Jifeng Yang and G-j Ni, Acta. Phys. Sinica, {\bf 4}, 88 
(1995); Jifeng Yang, Report No. hep-th/9801004.
\bibitem {JF2} G-j Ni and Jifeng Yang, Phys. Lett. {\bf B 393}, 79 (1997).
\bibitem {JH} see, e.g., J. H. Schwarz, Nucl. Phys. {\bf B} (Proc. Suppl.) 
{\bf 55 B},1 (1997); hep-th/9607201; or that given by W. Lerche, 
hep-th/9710246
\bibitem {Hie} G. 't Hooft, in {\it Recent Developments in Field Theories},
eds. G. 't Hooft et al. (Plenum Press, New York, 1980);
E. Witten, Nucl. Phys. {\bf B 188}, 513 (1981).
\bibitem {SUSY} D. Volkov and V. P. Akulov, Phys. Lett {\bf B 46}, 109 
(1973);
J.Wess and B. Zumino, Nucl. Phys. {\bf B 70}, 39 (1974).
\bibitem {WeinEFT}  S. Weinberg, Phys. Lett. {\bf B 251}, 288 (1990);
Nucl. Phys. {\bf B 363}, 1 (1991).
\bibitem {EFFT} For pedagogical reviews, see D. B. Kaplan, Report No. 
nucl-th/9506035; or A. V. Manohar, Report No. hep-ph/9606222.
\bibitem {Beg}  M.A.B. B\' eg and R.C. Furlong, Phys. Rev. {\bf D21},
1370 (1985).
\bibitem {Cohen} T.D. Cohen, Phys. Rev. {\bf C 55}, 67(1997).
\bibitem {PC} D.R. Phillips and T.D. Cohen, Phys. Lett.{\bf B 390}, 7 
(1997).
\bibitem {Jackiw} R. Jackiw, in {\it M. A. B. B\' eg Memorial Volume, 
} A. Ali and P. Hoodbhoy, eds. ( World Scientific, Singapore, 
1991), p25-42.
\bibitem {AG} see, e.g., N. Akhiezer and I.M. Glazman, {\it Theory of
Linear Operators in Hilbert space, } Ch. VIII. (Vishtcha Shkola, 
Kharkov, 1978); Or the book by M. Reed and B. Simon cited above.
\bibitem {Albeverio} A. Albeverio, {\it et al,  Solvable Models in 
Quantum Mechanics, } (Springer-Verlag, Berlin, 1988) and the 
extensive references therein.
\bibitem {Grossmann} A. Grossmann, R. Hoegh-Krohn and M. 
Mebkhout, J. Math. Phys. {\bf 21}, 2376 (1980).
\bibitem {GEP} G.-j. Ni,{ \it et al, } Report No. hep-ph/9801264.
\bibitem {DAI} X.-x. Dai, {\it et al }, Phys. Rev. {\bf A 55}, 2617 (1997) 
and references therein.
\bibitem {Case} K.M. Case, Phys. Rev. {\bf 80}, 797 (1950).
\bibitem {CPT} J. Gasser and H. Leutwyler, Ann. Phys. (NY) {\bf 158}, 
142 (1984); Nucl. Phys. {\bf 150}, 465 (1985).
\bibitem {Dono} J.F. Donoghue, Phys. Rev. {\bf D 50}, 3874 (1994).
\bibitem {Cornwall} J.M. Cornwall, R. Jackiw and E. Tomboulis, Phys. 
Rev. {\bf D 10}, 2428 (1974).
\bibitem {Kino} T. Kinoshita,  J. Math. Phys. {\bf 3}, 650 (1962); 
T. Kinoshita and A. Ukawa, Phys. Rev. {\bf D 13}, 1573 (1976).
\bibitem {PQ} E. C. Poggio and H. R. Quinn, Phys. Rev. {\bf D 14}, 
578 (1976).
\bibitem {Muta} T. Muta, {\it Foundations of Quantum chromodynamics, 
} World Scientific, Singapore, 1987.
\bibitem {LN} T. D. Lee and M. Nauenberg, Phys. Rev. {\bf B 133}, 
1549 (1964).
\bibitem {Haag} R. Haag, {\it Local Quantum Physics, }
Springer-Verlag, Berlin, 1993.
\bibitem {Confine} K. G. Wilson, Phys. Rev. {\bf D 10}, 2445 (1974);
A. Casher, J. Kogut and L. Susskind, Phys. Rev. {\bf D 10}, 732 (1974).
\bibitem {SW}  In the recent progresses in SUSY dual models \'a la
Wilsonian actions, the duality almost automatically defined the IR 
functional form of the LE theories. See, N. Seiberg and E. Witten, Nucl. 
Phys. {\bf B 426}, 19 (1994); (E) {\bf B430}, 485(1994);
Nucl. Phys. {\bf  B431}, 484(1995). For comprehensive reviews, see,
M. Shifman, Prog. Part. Nucl. Phys. {\bf 39}, 1 (1997) and M. Peskin,
hep-th/9702094. A pedagogical introduction has been given by A. Bilal,
hep-th/9601007.
\bibitem {Buch} D.B. Buchholz, Phys. Lett. {\bf B 174}, 331 (1986).
\bibitem {JY1} Jifeng Yang, in progress.
\bibitem {Measure} Recent efforts in regularizing the jacobians can be
found in F. De Jonghe, {\it et al, }, Phys. Lett. {\bf B 289}, 354 (1992) 
and in W. Troost, {\it et al, } Nucl. Phys. {\bf B 33}, 727 (1989) and the 
references therein.
\bibitem {JY2} Jifeng Yang, in progress.
\bibitem {BV} For comprehensive review on this subject, see, J. Gomis,
{\it et al, } Phys. Rep. {\bf 259}, 1 (1995).
\bibitem {gom} J. Gomis and S. Weinberg, Nucl. Phys. B {\bf 469}, 473 
(1996) and references therein.
\end{thebibliography}
\end{document}